\documentclass[conference]{IEEEtran}

\usepackage[utf8]{inputenc}
\usepackage{kotex}
\IEEEoverridecommandlockouts
\usepackage{cite}
\usepackage{amsmath,amssymb,amsfonts,comment, hyperref, diagbox}
\usepackage{algorithmic}
\usepackage{graphicx}
\usepackage{textcomp}
\usepackage{xcolor, fancyhdr}
\def\BibTeX{{\rm B\kern-.05em{\sc i\kern-.025em b}\kern-.08em
    T\kern-.1667em\lower.7ex\hbox{E}\kern-.125emX}}
\begin{document}

\title{Vocoder-free End-to-End Voice Conversion with Transformer Network
\thanks{Preprint. Work in progress}
}
\author{\IEEEauthorblockN{June-Woo Kim}
\IEEEauthorblockA{\textit{Department of Sensor and Display Engineering} \\
\textit{Kyungpook National University}\\
Daegu, Republic of Korea \\
kaen2891@gmail.com}
\and
\IEEEauthorblockN{Ho-Young Jung}
\IEEEauthorblockA{\textit{Department of Artificial Intelligence} \\
\textit{Kyungpook National University}\\
Daegu, Republic of Korea \\
hoyjung@knu.ac.kr}
\and
\IEEEauthorblockN{Minho Lee}
\IEEEauthorblockA{\textit{School of Electronics Engineering} \\
\textit{Kyungpook National University}\\
Daegu, Republic of Korea \\
mholee@gmail.com}
}

\maketitle
\begin{abstract}
Mel-frequency filter bank (MFB) based approaches have the advantage of learning speech compared to raw spectrum since MFB has less feature size. However, speech generator with MFB approaches require additional vocoder that needs a huge amount of computation expense for training process. The additional pre/post processing such as MFB and vocoder is not essential to convert real human speech to others. It is possible to only use the raw spectrum along with the phase to generate different style of voices with clear pronunciation. In this regard, we propose a fast and effective approach to convert realistic voices using raw spectrum in a parallel manner. Our transformer-based model architecture which does not have any CNN or RNN layers has shown the advantage of learning fast and solved the limitation of sequential computation of conventional RNN. In this paper, we introduce a vocoder-free end-to-end voice conversion method using transformer network. The presented conversion model can also be used in speaker adaptation for speech recognition. Our approach can convert the source voice to a target voice without using MFB and vocoder. We can get an adapted MFB for speech recognition by multiplying the converted magnitude with phase. We perform our voice conversion experiments on TIDIGITS dataset using the metrics such as naturalness, similarity, and clarity with mean opinion score, respectively.%\footnote{Code can be provided when the paper is accepted.}
\thanks

%\lfoot{Preprint. Work in progress.}

\end{abstract}

\begin{IEEEkeywords}
voice conversion, vocoder-free, transformer, spectrum, phase
\end{IEEEkeywords}

\section{Introduction}
Voice conversion has gained considerable attention in various industrial areas. In recently, encoder-decoder models built with recurrent neural networks (RNNs), such as long short-term memory (LSTM)\cite{hochreiter1997long}, bidirectional long-short term memory (BiLSTM)\cite{schuster1997bidirectional}, and gated recurrent unit (GRU)\cite{chung2014empirical} have been widely utilized for sequence modelling. There are lots of neural network models based on RNN encoder-decoder structure also known as sequence-to-sequence (Seq2Seq)\cite{sutskever2014sequence} and they achieved good results for voice conversion tasks.

RNNs, however, process words one by one for each sequence. This sequential property of RNNs can be an obstacle for parallel computation of GPUs and make training slow. On top of that, if these temporal information gets longer, the model tends to forget the contents of distant locations or mixes it with the contents of the next location. The transformer\cite{vaswani2017attention} network partially solved these problems of RNNs by using an attention mechanism to derive global dependency between input and output, which reached state-of-the-art performance in many fields. Transformer which does not have any convolutional neural network (CNN)\cite{kim2014convolutional} or RNN layers has shown the advantage of learning fast and solved the limitation of sequential computation of conventional RNN.

Given the waveform speech as the model input for voice conversion, the short-time Fourier transform (STFT) converts it into raw spectrum in time-frequency domain form. This spectrum computed with STFT can provide useful information than waveform speech. In particular, the conventional approaches used in text-to-speech (TTS), voice conversion, and speech recognition pass Mel filter banks through raw spectrum to generate Mel-frequency filter bank (MFB, also called Mel-spectrogram). In MFB, the frequency components of the spectrum are obtained after STFT. After that, it is compressed according to the Mel curve\cite{logan2000mel} reflecting the characteristics of the Cochlea in the human ear. In MFB, phase information is removed when it is compressed.

MFB, which consist of 40 to 80 feature dimensions per time step, has the advantage of learning speed compared to raw spectrum since MFB has less feature size. However, it can't be converted directly to waveform speech because of losing phase information. Thus, speech generator with MFB approaches require additional vocoder that needs a huge amount of computation expense for training process. In other words, MFB fed into the Seq2Seq should be synthesized for natural speech through phase estimation with the help of vocoder, which synthesizes the linear scale spectrum. Then, it can get the final output of the model into waveform speech.

Thus, speech generator with MFB approach requires additional vocoder that demands a heavy training computation process. Using the vocoder such as Griffin-Lim\cite{griffin1984signal} and WaveNet\cite{oord2016wavenet}, it is possible to get better quality of voice with the synthesis. On the contrary, it is inevitable to avoid problem with complexity due to the extra computation.

However, to avoid additional pre/post processing such as MFB and vocoder, we propose a fast and effective approach to convert realistic voices using raw spectrum in a parallel manner to generate different style of voices with clear pronunciation. In this paper, we introduce a vocoder-free end-to-end voice conversion method using transformer network. We focus on the converting the raw spectrum obtained by the STFT without help of the vocoder which requires iterative synthesis. In addition, it is possible to use phase information to restore the waveform speech through inverse STFT.

Our presented conversion model can also be used in speaker adaptation for speech recognition. Our approach can convert the source voice to a target voice without using MFB and vocoder. We can get an adapted MFB for speech recognition by multiplying the converted magnitude with phase. Furthermore, it is also possible to convert voices of minorities (elderly, children, dialects, speaker with disabilities) with poor speech recognition performance to those of common adults. It is possible to achieve better speech recognition performance through speaker adaptation which replaces the features of minorities and common adults. We perform our voice conversion experiments on TIDIGITS dataset using the metrics such as naturalness, similarity, and clarity with Mean Opinion Score (MOS), respectively.

\section{Related Work}
In this section, we first introduce the prior research on vocoder, voice conversion, and the transformer network that we used in this paper.
\subsection{Vocoder}
Vocoder is used to synthesize linear scale spectrum into speech signals, synthesizing natural speech through phase estimation. In Griffin-Lim algorithm \cite{griffin1984signal}, the STFT of the speech signal output in the previous step is calculated and the amplitude is replaced by the modified-STFT magnitude given as input. This algorithm recovers speech signals with the STFT magnitude which is the most similar to a given modified-STFT through an iterative process of restoring the original signal to minimize the squared error of the amplitude of the new STFT and the input modified STFT.

WaveNet\cite{oord2016wavenet} is an autoregressive model that uses sequential features between speech samples and has succeeded in synthesizing high quality speech by predicting the next sample using previous samples. However, the speed of the generation rate is slow because the next sample is generated one by one from the previous samples. Parallel WaveNet\cite{oord2017parallel} is designed to solve the WaveNet's slow sample generation speed, which uses inverse autoregressive flow to synthesize voices. Since inverse autoregressive flow does not know the distribution of the target voice data set during learning, the learning is performed by extracting the distribution information of the target data set using a well-trained WaveNet (teacher network) and comparing it with the result of inverse autoregressive flow. It has the advantage of faster speech synthesis than WaveNet, but the quality of synthesized speech is lower. Unlike \cite{oord2017parallel}, WaveGlow\cite{prenger2019waveglow} is not required for a pre-trained teacher network and has the advantage of fast voice synthesis. However, since WaveGlow uses a distribution based loss function, the quality of synthesized speech is poor. Furthermore, when combined with TTS, poses the problem that the quality of synthesized speech depends on the quality of the predicted MFB from the text.

\subsection{Voice Conversion}
In \cite{biadsy2019parrotron}, the voices of speaker with a disability are converted into general voices. The encoder consists of CNN and three BiLSTMs, and the decoder consists of two LSTMs. Attention between encoder-decoder is used. In order to solve the problem of signal-to-signal conversion, the speech recognition decoder is connected to the encoder output for the multitask learning\cite{caruana1997multitask} and it used only in the training task. 

To translate between voices of different languages and synthesize the translated output as speech, usually it had to go though speech recognition, translation, and TTS tasks. In this paper\cite{jia2019direct}, however, they convert the speech of different languages into an end-to-end attention-based Seq2Seq network. Without going through other steps, it can directly translate the speech of another language into one. Encoder is composed of 8 BiLSTMs, and the encoder output is used to predict the phoneme temporal information of input and target through auxiliary tasks. Likewise\cite{biadsy2019parrotron}, these auxiliary decoders were used only for learning. In addition, the decoder can be optionally adjusted according to the input speaker. Thus, voice can be converted to the desired speaker's voice by using pre-trained speaker encoder. They consider to use WaveRNN vocoder\cite{kalchbrenner2018efficient} rather than Griffin-Lim because it dramatically improves voice quality.

\subsection{Transformer network}
RNN is widely used method for sequence modeling tasks such as neural translation and language modeling. However, because the RNNs process words one by one for each sequence, this sequential process can be an obstacle with parallelization and slow learning. On top of that, if these temporal information get longer, the model tends to forget the contents of distant locations in order or to mix with the contents of the next location.
\begin{figure}[htbp]
    \centering
    \includegraphics[scale=0.5]{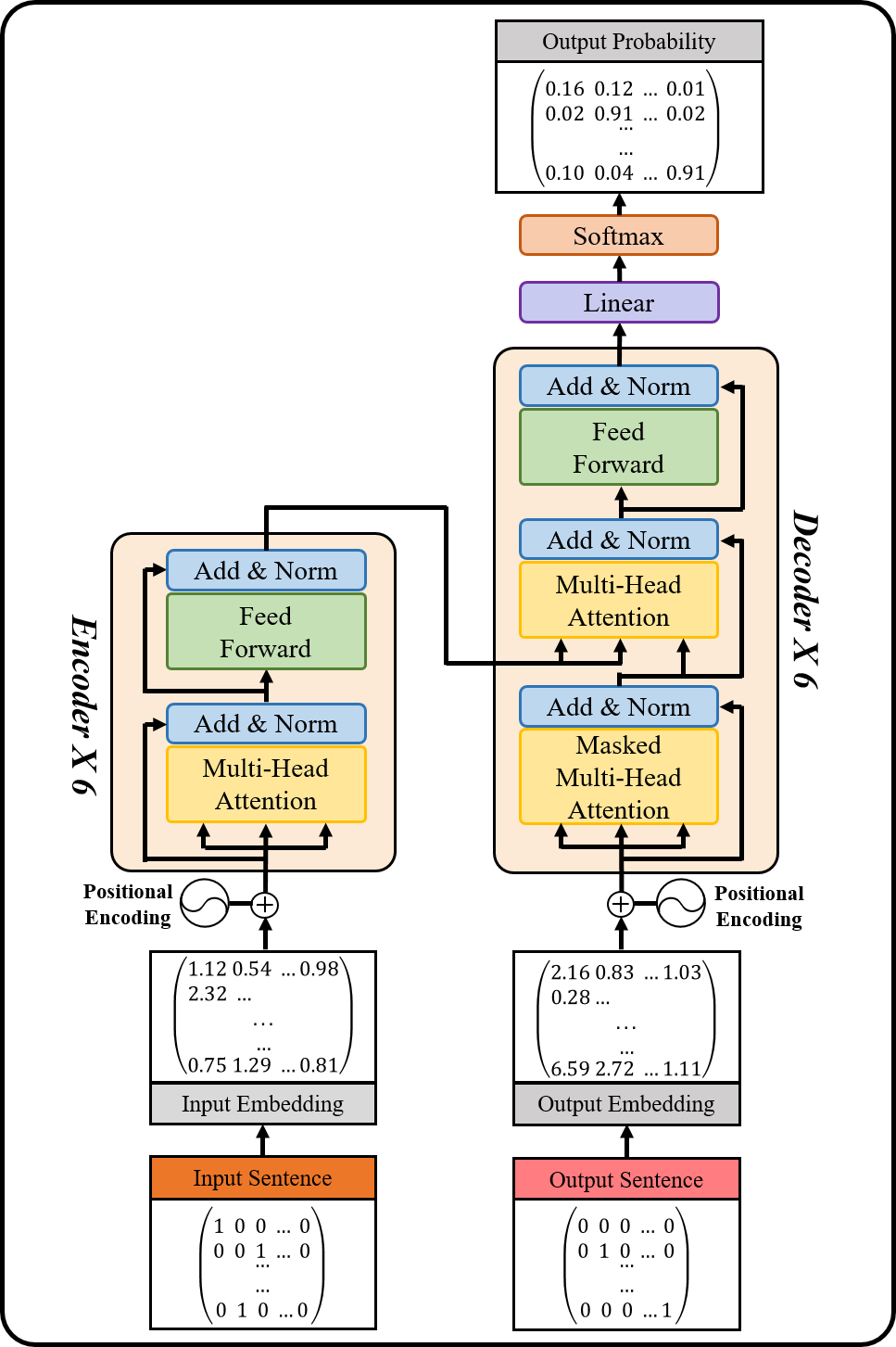}
    \caption{Vanilla transformer network.}
    \label{fig:transformer_original}
\end{figure}
The transformer network in \cite{vaswani2017attention}, is the model architecture that relies entirely on attention mechanisms to derive global dependencies between inputs and outputs. As Fig. \ref{fig:transformer_original} shows, the transformer model architecture without CNN and RNN have shown the advantage of fast learning time. The shortcomings of traditional RNN due to poor performance in temporal information, have been solved with self-attention. BERT\cite{devlin2018bert}, which is evolved from transformer, is used in many natural language processing (NLP) fields such as not only translation but also summary and prediction of sentence relevance, etc. BERT is widely used in other fields along with NLP. VideoBERT\cite{sun2019videobert} learned two-way joint distribution of visual and linguistic token sequences derived from vector bidirectional and speech recognition with output of video data. This has led to the research in a variety of tasks, including action classification and video captions. In \cite{li2019neural}, combination transformer network with TTS model which is called Tacotron2\cite{shen2018natural}, used to present the results of speech synthesis. In \cite{huang2019voice}, which performs voice conversion based on the transformer network, uses pre-trained TTS. They perform voice transformation with pre-trained model parameters using vocoder based synthesis.

Consequently, vocoder helps to improve the quality of speech synthesis, but it takes time to synthesize. We use transformer network due to its generalization performance through self-attention as well as fast and effective parallel learning techniques. In addition, we perform our experiments by focusing on the conversion of raw spectrum stage without adopting the voice synthesis method through the vocoder. More details are given in section 3.

\section{Method}
This section introduces using raw spectrum rather than MFB for end-to-end voice conversion without the help of vocoder.
\subsection{Raw spectrum}
\begin{figure}[h]
    \centering
    \includegraphics[scale=0.5]{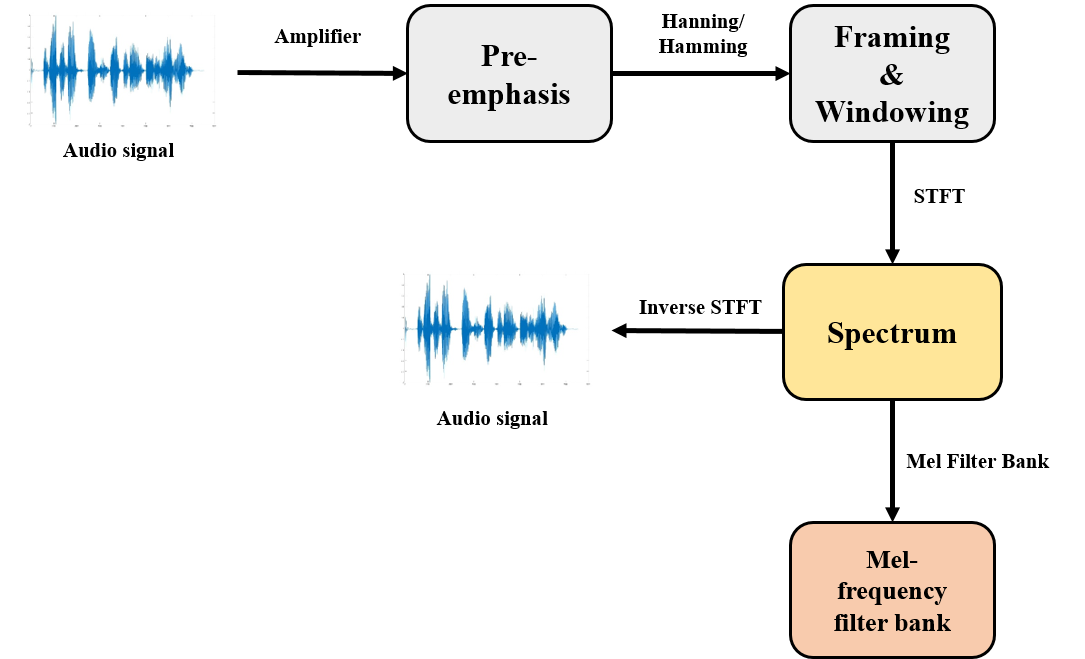}
    \caption{A series of steps to get spectrum and MFB from audio signal.}
    \label{fig:fft}
\end{figure}
Fig. \ref{fig:fft} shows a flowchart that converts waveform speech into spectrum, MFB, and back to waveform speech. Given a continuous audio signal $x[n]$, this can be expressed as:
\begin{figure*}[h!]
    \centering
    \includegraphics[scale=0.5]{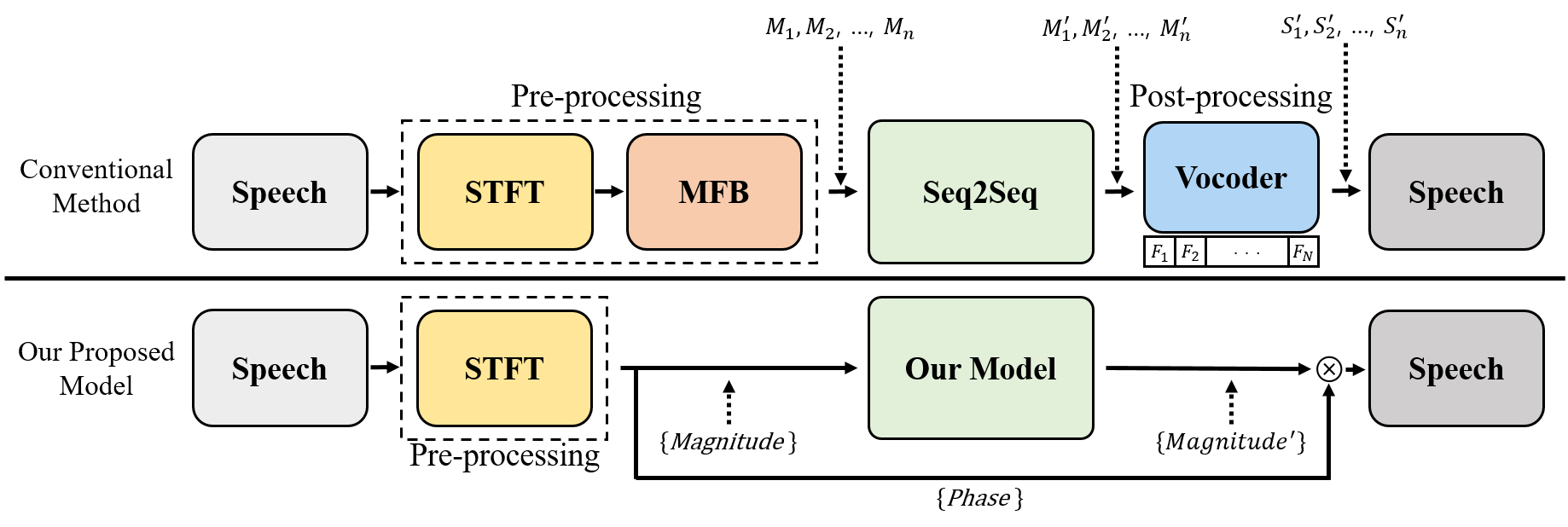}
    \caption{Difference between conventional method and our proposed model on voice conversion. The top part (Conventional Method), requires pre/post processing with MFB, but our proposed model only requires raw spectrum to make audio}
    \label{fig:Differencial from tranditional}
\end{figure*}
\begin{equation}
    x[n] = Acos(\omega nT+\phi) = Acos(2\pi f nT+\phi)\label{raw_spec1}
\end{equation}
where $A$ is amplitude, $\omega$ is angular frequency in radians/seconds, $f$ is $\omega/2\pi$, $\phi$ is initial phase in radian, $n$ is time index, and $T$ is $1\over{f_s}$, respectively. Next process is applying a pre-emphasis filter on the $x$ to amplify the high-frequency. Pre-emphasis filter is useful in several ways. High-frequency is generally smaller than low-frequency. Thus, using pre-emphasis filter helps to avoid numerical problems during STFT and improves signal-to-ratio.

As the frequency of the signal changes over time, after pre-emphasis, the signal is split into short time frames. Because of the frequency contour of the signal is lost over time, the Fourier transform is performed assuming that the frequency of the signal is stationary for a very short period of time, not over the entire signal. The typical frame size for speech processing is 20ms to 40ms, with 50\% overlap. Usually 25ms is used for frame size and 10ms (15ms overlap) for stride overlap size. 

The next step is to cut the signal into frames and apply the hamming, hanning window function to each frame. The spectrum can be calculated by performing an N-point FFT (NFFT) on each frame. Here, NFFT generally uses 256 ($16ms$) or 512 ($32ms$). Finally, the spectrum that is obtained through STFT can be expressed with magnitude and phase by the following equation:
\begin{equation}
    D = S * P\label{intro_eq1}
\end{equation}
where $D$ is complex-valued spectrum, $S$ is magnitude and $P$ is phase, respectively. 

In summary, raw spectrum can be recovered from speech waveform directly as shown in Fig. \ref{fig:fft}. Thus, we use spectrum to perform voice transformation in an effective way with out any post-processing.

\subsection{Proposed model structure}
\subsubsection{Model flow}
The vocoders mentioned in section 2 are complex and computationally expensive which require a lot of repetitive works to restore the audio waveforms. To solve this problem, we focus on the conversion at the spectrum level. Fig. \ref{fig:Differencial from tranditional} shows,  the conventional method of using MFB in the upper part, and the proposed transformer network in the lower part. 
%\cite{wang2017tacotron}
%In upper part in Fig. \ref{fig:Differencial from tranditional}, the MFB is used as input to the Seq2Seq model, and the output is obtained through the vocoder. 
%This requires both pre/post processing steps, which will ultimately take additional time to train and inference the model. 
Conventional method\cite{wang2017tacotron} uses the output of MFB expressed as $M_1, M_2, ..., M_n$ as input to Seq2Seq and the output is obtained through the vocoder. The encoder input in the Seq2Seq considers all the temporal information. It's no different from our model. However, the decoder predicts $n$ frames of MFB at once, thereby reducing the number of decoder steps to $n/\gamma$, where $\gamma$ is the reduction factor. Post-processing of linear scale spectrum $F$ is performed using CBHG (1D convolution bank, highway network, bidirectional gated recurrent unit) module which results in $F_1, F_2, ..., F_n$. Vocoder is essential to convert $F$ into a waveform expressed as $S_1', S_2', ..., S_n'$. Autoregressive vocoder which uses the previous input to predict the current step is used. Once we get $S_1'$, use $S_1'$ to predict $S_2'$ and finally $S_n'$ is obtained. However, this cannot reduce the computation cost.

On the other hand, in the proposed model in Fig. \ref{fig:Differencial from tranditional}, the  magnitude $S$ and phase $P$ are obtained using Eq. \eqref{intro_eq1} from the raw spectrum after passing through STFT. Then, we use $S$ as input to the model encoder. The decoder converts the spectrum in parallel. After element-wise multiplication between final output of the model $\hat{x}$ and input phase $P$, it is possible to get a converted target speech by inverse STFT. We can recover the predicted voice instantly using the converted magnitude and phase of the source without help of the vocoder. Our proposed model is a fast and effective approach to convert realistic voices using raw spectrum in a parallel manner. Our method does not dependent on post-processing. 

\subsubsection{Tokens and zero-padding}
The model input of the corpus is usually vectorized from word embedding matrix. The spectrum, unlike corpus, consists of continuous values. The spectrum consists of $N$ dimensions by time $T$. These values are not sparse representations. Corpus sets the maximum length and proceeds with a start of sentence (SOS) into in front of corpus and the end of sentence (EOS) at the end of corpus.

The SOS that combined sequence is used as the decoder input. Seq2Seq needs to be trained with real values by teacher forcing. Thus, in the inference phase, the input of decoder uses only SOS token.

Through this, the autoregressive transformer performs prediction using beam search or greedy search. We put the EOS token into our decoder input and perform voice conversion. In addition, since beam search is based on beam depth and softmax, we use greedy search.

We used zero-padding for all the spectrum. The reason for using zero-padding is that the transformer network considers the whole sequence and learns in parallel. Even if the voice scripts are the same, the length of each speaker's characteristics is different. For this reason we used zero-padding. 

In order to avoid attention between the zero value and the real vector, we multiplied $-1e-9$ when there is a zero value on dimension in each time step. Zero-padding is described in the next section.
%왜 우리가 패딩을 사용했는지, SOS EOS 왜 안되는지, (256,1)에 의한, 비슷한 값, 등등 논리적으로 나열, 내용 추가 예정

\subsubsection{Transformer-based model architecture}

\begin{figure}[h!]
    \centering
    \includegraphics[scale=0.55]{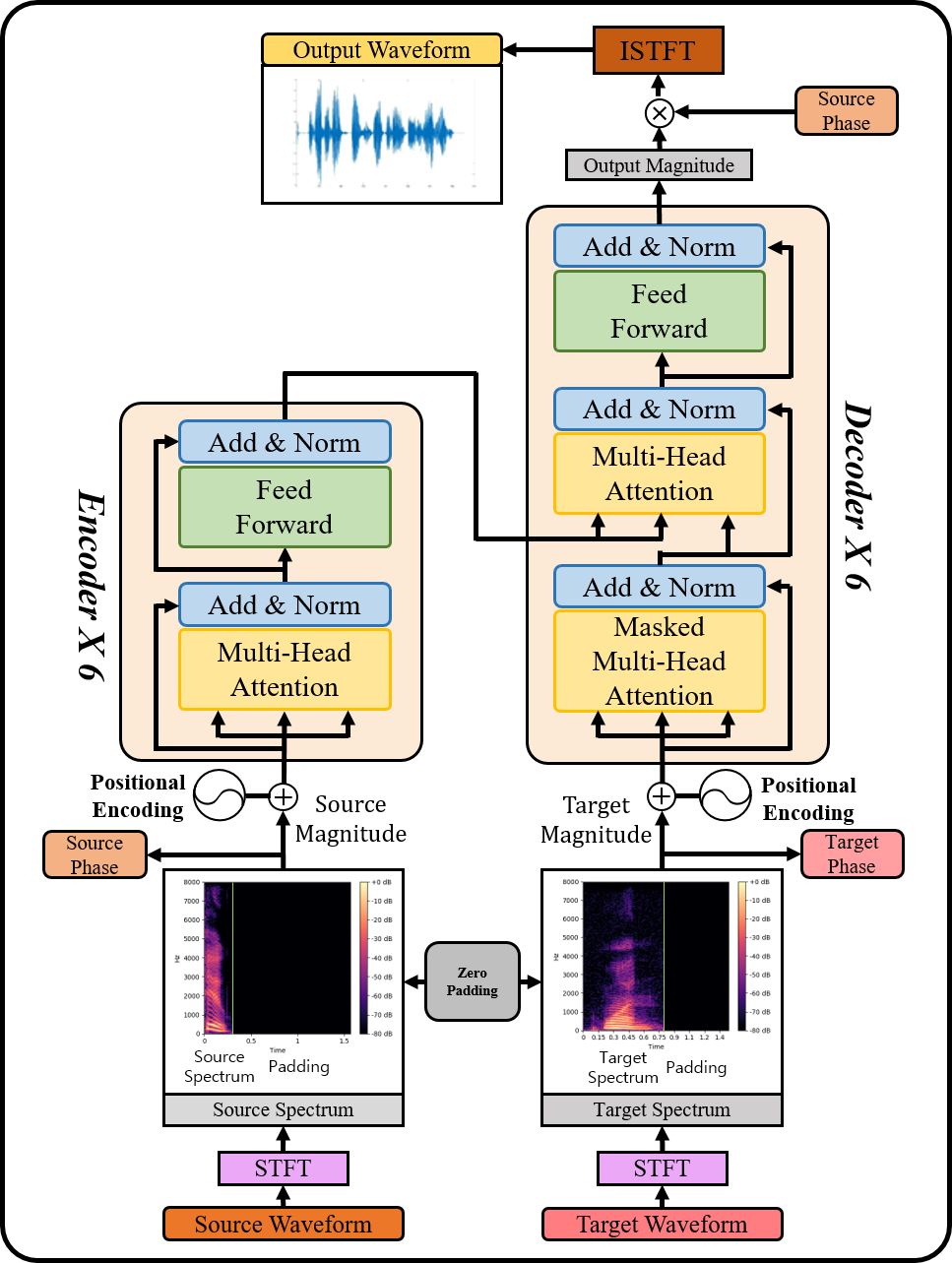}
    \caption{Our transformer-based model architecture. The input of the encoder is magnitude of raw spectrum and the output of decoder is converted magnitude. Predicted x perform element-wise multiplication with source phase. Word embedding, output linear, and softmax function are not needed, respectively.}
    \label{fig:our_transformer}
\end{figure}

Fig. \ref{fig:our_transformer} shows our transformer-based model architecture. 
%%%%%%%%%%%%%%%%%%%%%%%%%%%%%%%%%%%%%%%%%%%%%%
Firstly, we obtain a spectrum that depends on the $NFFT$ coefficients and then separate $S$ and $P$ by Eq. \eqref{intro_eq1}. After that, $S$ is used as the encoder input. In this case, we don't use word embedding\cite{levy2014neural} because the $S$ is a time-frequency domain that consists of sampling the frequency along the time axis. The final input is $S$ plus the position vector passed through PE. Then multi-head attention is performed through $N$-encoders. The multi-head attention results pass through two-layers feed-forward network that contain rectified linear units (ReLU)\cite{nair2010rectified} to reconstruct information which are not cleaned up. The process up to now is to make a new context information by combining the entire temporal information for each time step. Then we perform a residual connection\cite{he2016deep} that adds input data to the values which are obtained until now. This means that context information which are not included in the input temporal information are processed by the input and added. The encoder looks at the entire given temporal information and encodes each time step information into a better representation.

%Fig. \ref{fig:1} shows our model architecture. $NFFT$ 계수에 의존한 spectrum을 얻고 ``\eqref{intro_eq1}'' 에 의한 $S$와 $P$를 분리합니다. 이후, Encoder input으로 magnitude가 사용됩니다. 이 때, spectrogram은 time 축으로 주파수를 샘플링하여 이루어진 주파수/시간 도메인이기 때문에 우리는 Word Embedding\cite{levy2014neural}을 사용하지 않습니다. 입력 $S$에 PE를 통과한 위치 벡터가 더해진 값이 최종 input이 됩니다. 그 후 $N$개의 encoder를 통해 Multi-Head Attention (MHA)을 행하고, MHA 과정만으로 정리되지 않는 정보를 재정리 하는 역할을 Rectified Linear Units (ReLU)\cite{nair2010rectified}가 포함된 2 layers Feed-forward network (FFN)에서 하기 위해 통과합니다. 현재까지의 과정은 각 time step별로 sequence 전체 정보들을 종합하여 새로운 context 정보를 만드는 역할이었습니다. 현재까지 구해진 값을 input data와 더하는 Residual Connection \cite{he2016deep}을 합니다. 이는 input sequence에 포함되지 않는 context information을 input sequence로부터 가공하여 더한다는 의미입니다. encoder는 주어진 sequence 전체를 살펴보며 각 time step information을 더 나은 representation으로 encoding 하는 역할을 합니다. 

Decoder uses only $S$ which passed through STFT from spectrum signal of target $y$ like the encoder method, and creates new information based on the known information. However, the decoder is different that it uses masked multi-head attention when performing self-attention. The reason for using masked multi-head attention is to prevent self-attention by covering features after its time step during self-attention. This shows the transformer network is autoregressive model. After that, attention is concatenated between the encoder outputs and decoder outputs. This process determines how much decoder uses $x$ of input spectrum temporal information to express $y_i$. The results of encoder-decoder attention are added to the masked multi-head attention results of the decoder. Then they are put into the feed forward network. The outputs finally come out. So far, the outputs $\hat{x}$ have the same dimension $d_{model}$ as inputs $x$ and targets $y$, only the magnitude temporal lengths are different. Finally, $\hat{x}$ currently only have magnitude that converted from source $x$ to target $y$. Then we multiply this values by $P$ to make a spectrum containing complex numbers. Finally, it can be restored to waveform speech using inverse STFT.

%Decoder에서는 위의 방식과 마찬가지로 target $y$의 파형 시그널에서 STFT를 통과한 $S$만 사용하며, 현재까지 알려진 정보를 바탕으로 새로운 정보를 생성합니다. Encoder와 동일하지만, Self-Attention시 Masked-Multi-Head Attention (Masked-MHA) 을 쓴다는 점이 다릅니다. Masked를 사용하는 이유는 self-Attention시 자신의 time step 이후의 feature를 가려 self-attention이 이루어지는 것을 막는 역할을 합니다. 그 후 encoder-decoder attention이 이루어지는데, decoder가 $y_i$의 정보를 표현하기 위하여 input sequence의 $x$의 정보를 얼마나 이용할지 결정하는 역할을 합니다. 최종적으로 decoder의 Maksed-MHA 결과에 encoder-decoder attention의 결과가 더해져서 FFN에 입력이 되고, 최종 출력이 나오게 됩니다. 현재까지의 출력값 $\hat{x}$의 dimension은 input x와 target y와의 $d_{model}$이 같고, sequence length만 다릅니다. 그렇기에 $\hat{x}$에는 현재 변환된 target의 magnitude만 존재하고 있으며, 이 값에 $P$를 곱하여 복소수가 포함된 spectrogram으로 만듭니다. 최종적으로 ISTFT를 거쳐서 파형 스피치로 복원합니다.

The transformer has fewer parameter numbers than other models, and because it uses feed forward network, parallelism is easy and fast operation is possible. Nevertheless, accurate modeling is possible because information between distant temporal information are directly linked.
%transformer는 다른 모델들보다 parameter 숫자가 적고, FFN을 사용하기 때문에 병렬화가 쉽고 빠른 연산이 가능합니다. 그럼에도 불구하고 멀리 떨어진 sequence들 간의 정보가 곧 바로 연결되기 때문에 정확한 모델링도 가능합니다.

\section{Experimental setup}~\label{sec:experiments}
In this section, we introduce the dataset, pre-processing, and hyperparameters.
\subsection{Database and feature extraction}
We use the TIDIGITS\cite{tidigits} dataset which consists of 326 speakers (111 men, 114 women, 50 boys, 51 girls) pronouncing numbers. Among them we experiment with independent numeric units (e.g., "one", "two", ..., "oh", "zero"). Our experiments require the pair of source and target dataset from each different speakers. Therefore, we train a paired dataset of 55 men, 57 woman, 25 boys and 26 girls. There are two numerical data of each train dataset. Because of TIDIGITS test dataset was separated, we use as it is. The sampling rate of the corpus is $20\;kHZ$ and dataset was collected with an Electro-Voice RE-16 Dynamic Cardiod microphone in a quiet space.

%우리는 TIDGITS이라는 숫자를 발음하는 326 명의 speaker (111 men, 114 women, 50 boys, 51 girls)로 이루어진 데이터셋을 사용합니다. 그 중 우리는 독립 숫자 단위에 대해 실험을 진행하였습니다. (e.g., "one", "two", ..., "oh", "zero") 이 중 paired dataset인 55 men, 57 woman, 25 boys, 26 girls 에 대해 train을 하였고, test는 TIDGITS에서 나눠 놓은 train에서의 unseen test set에 대하여 진행하였습니다. corpus의 sampling rate는 20kHZ이며 조용한 공간에서 Electro-Voice RE-16 Dynamic Cardiod microphone으로 수집되었습니다.

We downsampled $20\;kHZ$ to $16\;kHZ$ for reducing the computation. We pre-processed the dataset such as $NFFT$ is 512 ($32ms$) and $hop\_length$ is 256 ($16ms$) to get raw spectrum. The dimension of the obtained spectrum is $(257,T)$. However, since the transformer $d_{model}$ is $2^n$, we intentionally remove the last imaginary part of the spectrum. 
%우리는 계산량을 줄이기 위해 16kHZ로 downsample하여 진행하였습니다. raw spectrogram을 뽑기 위한 STFT에 의한 계수로 $NFFT$는 512 ($32ms$)를 사용하였고, $hop\_length$는 256 ($16ms$)입니다. 얻어진 spectrogram의 dimension은 $(257,T)$가 됩니다. 하지만 transformer의 $d_{model}$이 $2^n$이기 때문에, 우리는 spectrogram의 가장 마지막인 imaginary part를 의도적으로 제거하여 $(256,T)$를 맞춘 상태에서 model의 input으로 사용하였습니다.

\subsection{Data pre-processing}
\subsubsection{Voice Activity Detection}
\begin{figure}[h!]
    \centering
    \includegraphics[scale=0.5]{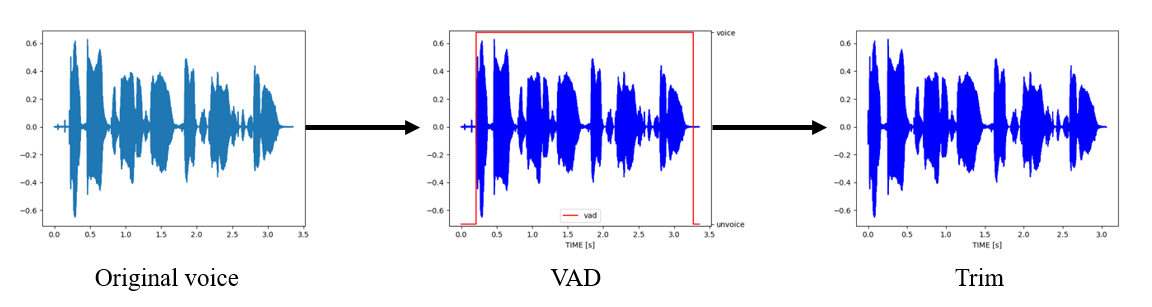}
    \caption{Original wav (left), VAD (middle), trimmed (right).}
    \label{fig:vad}
\end{figure}
Voice Activity Detection (VAD) is a technology applied to voice processing that detects the presence or absence of human voice. As shown in Fig. \ref{fig:vad}, VAD is an algorithm that determines the threshold criteria which is used to distinguishes background noise from real speech mainly used in speech recognition. We use VAD to reduce the maximum sequence length of the dataset by removing the front and back silence sections based on threshold to speed up computation. Through the pre-processing, this technique not only make our model accelerate learning but also prevent the learning to be difficult as the magnitude temporal information get longer.
%Voice Activity Detection (VAD)는 사람 음성의 유/무를 검출하는 음성 처리에 적용되는 기술입니다. Fig. \ref{fig:vad}에서 볼 수 있듯이, VAD는 배경잡음과 실제 음성(소리)를 구분하는 threshold 기준을 결정하는 알고리즘입니다. 주로 음성 인식에서 많이 사용되는 method이며, 우리는 VAD를 통해 앞 뒤의 silence 구간을 제거함으로써 dataset의 max sequence length를 줄여서 계산량을 빠르게 하여 모델의 학습이 빠르게 가속화 할 수 있도록 도움을 줌과 동시에 sequence가 길어짐에 따라 학습이 잘 되지 않는 것을 방지시켰습니다. 
\subsubsection{SOS token, EOS token, Padding}
In natural language processing, the first token of a sentence is start-of-sentence (SOS), and the end token is end-of-sentence (EOS). Usually, EOS is used because let model to know the input sentence is over. In addition, SOS token is utilized in inference phase as decoder input. Thus, we create SOS and EOS token corresponding to the $(256, 1)$ dimension which are randomly uniform distribution between 0 and 1. We concatenate the SOS token in front of the decoder inputs in all train dataset. Moreover, we use zero-padding at the end after concatenating the EOS token with target dataset. In the test phase, we put SOS token into the decoder and our model inference prediction using greedy search. 

In the last part of pre-processing is padding. We find the maximum sequence lengths in the train dataset. In order to match the same magnitude temporal information, we use zero-padding with all train dataset to maximum sequence lengths. In model training, $-1e-9$ values are used to prevent multi-head attention from occurring in the zero-padding part.

%NLP에서 문장의 첫 부분은 start-of-sentence (SOS), 문장의 끝에 End-of-sentence (EOS)를 넣기도 합니다. 문장이 끝났다는 것 까지 학습하도록 하기 위해 EOS를 넣습니다. 또한 inference phase에서 target의 값을 주지 않고 SOS token만 넣어주기 위해 우리는 EOS와 SOS token을 0부터 1 사이의 random uniform한 $(256, 1)$ dimension에 해당하는 값을 만들었습니다. 모든 train dataset내의 target 값 앞쪽에 SOS token을 concatenate 하였고, 마지막 부분에 EOS token을 concatenation한 뒤 zeropadding을 하였습니다. Inference 시에는 위와 같은 SOS token을 사용하여 decoder input으로 넣어 greedy search를 사용하여 추론하였습니다. 마지막으로 같은 sequence로 맞춰주기 위해 train set의 max sequence length를 구하여 zeropadding을 하였고, model 학습시에는 zero 구간에 대해 MHA가 일어나지 않도록 $-1e-9$값을 취하였습니다.
\subsection{Hyperparameter}
We use the Adam optimizer \cite{kingma2014adam} with $\beta_1=0.9$, $\beta_2=0.98$, and $\epsilon=1e-9$, respectively. Since the number of train dataset is small, we cannot adopt the original learning rate in~\cite{vaswani2017attention}. On the other hand, our initial learning rate is at $1e-4$ with proceeds to $decay\_step$ is 4000 and $decay\_rate$ is $0.96$.

We implement our model with Tensorflow 2.0 and train with one Titan RTX GPU. However, since we have small paired dataset and no post-processing, it's enough to use one 1080TI GPU in our experiments. In inference phase, the testing GPU memory that we only used is around 500 to 550MiB.
%우리는 $\beta_1=0.9$, $\beta_2=0.98$ and $\epsilon=1e-9$인 Adam optimizer\cite{kingma2014adam}를 사용하였습니다. 우리는 \cite{vaswani2017attention}와 다르게 dataset의 수가 작기 때문에 learning rate는 $1e-4$에서 시작하여 $decay\_step=4000$, $decay\_rate=0.96$으로 진행하였습니다.

\begin{table}[htbp]
\caption{Model Hyperparameters}
\begin{center}
\begin{tabular}{|l|r|}
\hline
\textbf{Hyperparameters}&\textbf{Value} \\
\cline{1-2}
\hline
      \hline
      $N_{encoder}$ & $6$ \\
      \hline
      $N_{decoder}$ & $6$ \\
      \hline
      $N_{heads}$ & $8$ \\
      \hline
      $d_{model}$ & $256$ \\
      \hline
      $d_{ff}$ & $1,024$ \\
      \hline
      $D_{rate}$ & $0.1$ \\
      \hline
\end{tabular}
\label{tab1}
\end{center}
\end{table}
Table~\ref{tab1} shows the hyperparameters. Six encoders and decoders, eight multi-head attentions were used in our model. The model size $d_{model}$ is 256 and the dimension size used for feed forward network $d_{ff}$ is 1024. Dropout \cite{srivastava2014dropout} selected 0.1 and it was used for training only. We adopt two losses.
%Table \ref{tab1} 은 우리가 사용한 hyperparameters에 대해 보여줍니다. encoder와 decoder는 각각 6개, MHA는 8개를 사용하였습니다. 모델 사이즈는 256으로 진행하였고, FFN에 쓰이는 dimension 크기는 1024입니다. Dropout\cite{srivastava2014dropout}은 0.1을 선택하였고 training에만 사용되었습니다.
%{\exp{(e_{ij)}}\over{\sum_{k=1}^{T_x}\exp{(e_{ik)}}}}\label{eq2}
%우리는 2개의 loss를 더하여 사용하였습니다.
\begin{equation}
    L_1 = \sum_{i=1}^n|y_{true}-y_{predicted}|\label{l1}
\end{equation}
\begin{equation}
    L_{MSE} = {1 \over 2}\sum_{i=1}^n(y_{true}-y_{predicted})^2\label{mse}
\end{equation}
\begin{equation}
    L_{final} = L_1*0.5 + L_{MSE}*0.5\label{l_final}
\end{equation}
Eq. \eqref{mse} has the advantage of minimizing the difference between variance and bias quickly, and Eq. \eqref{l1} tends to ignore the outlier, which is disadvantage of Eq. \eqref{mse}. Therefore, we use half of these equations under the hypothesis that they could complement each other.
%``\eqref{mse}''는 variance와 bias의 차이를 빠르게 최소화 하는 장점이 있고, ``\eqref{l1}은 ``\eqref{mse}''의 단점인 outlier를 무시하는 경향이 있기 때문에 서로 보완할 수 있다는 가설 아래에 절반씩 더하여 사용하였습니다. 

\section{Results}
In this section, we perform our voice conversion experiments on the TIDIGITS dataset using the metrics such as naturalness, similarity, and clarity with MOS.
%In this section, we display the voice conversion results on spectrum domain and provide our experiment results on TIDIGITS dataset that include the voice conversion results with gender differences in adults and children in terms of clarity and similarity assessments through MOS.

\begin{figure}[htbp]
    \centering
    \includegraphics[scale=0.5]{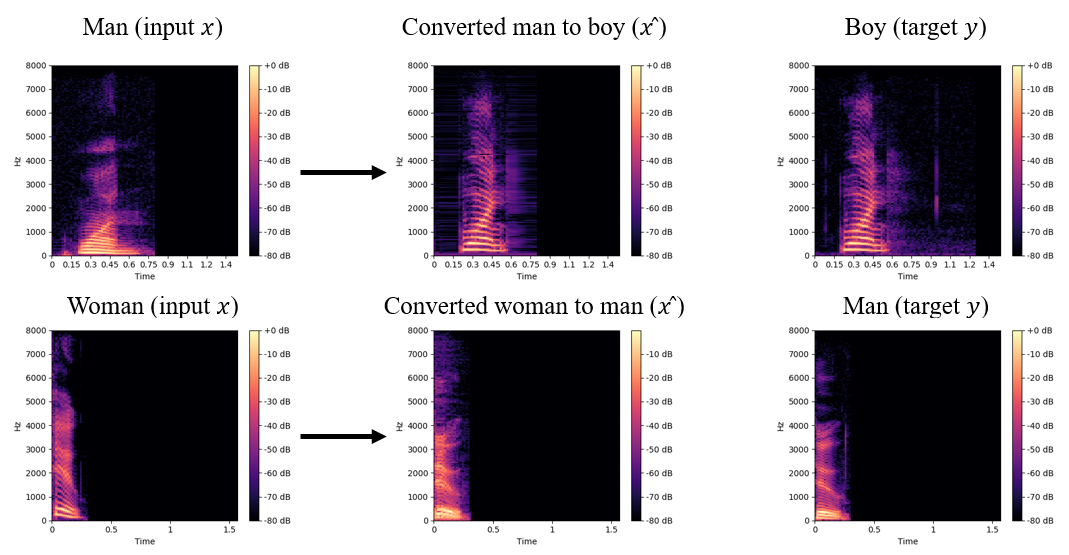}
    \caption{Visualization of our model conversion results. The figures (first row) show the inference results of conversion from man to boy with saying "1". The others (second row) show the results inference of conversion from woman to man with saying "5". In all figures, 8 kHz is maximum frequency corresponding to the y-axis}
    \label{fig:spectrogram_results}
\end{figure}
Fig. \ref{fig:spectrogram_results} shows the speech conversion results of our proposed model. The figures in the first row are the result of conversion from man to boy. The left figure in first row is the input of man, and the converted output is middle, and the source of boy is right in the figure, respectively. From the frequency in Fig.\ref{fig:spectrogram_results} on upper side, $\hat{x}$ period got wider than $x$ in 0 to $2\;kHz$ frequencies. In addition, part from $2\;kHz$ to $5\;kHz$ of $x$ does not have high dB, but the result of $\hat{x}$ is similar to $y$. Likewise, the input frequencies of second row of $x$ is converted similarly to $y$.
%Fig. \ref{fig:spectrogram_results}는 우리가 제안한 모델의 음성 변환 결과입니다. First row에 있는 figure들은 men의 음성이 boy 음성으로 변환된 결과입니다. 왼쪽이 input으로 들어가서 가운대 figure로 output된 것이며, $\hat{x}$의 0부터 2kHz frequency들의 주기가 $x$보다 넓어졌습니다. 또한 $x$의 2kHz부터 5kHz까지는 주황색으로 표시된  높은 dB들이 존재 하지 않지만, $\hat{x}$의 결과에선 $y$와 유사하게 존재하는 것을 볼 수 있습니다. 마찬가지로 Second row의 input $x$의 frequency들도 $y$처럼 유사하게 변환되었음을 보여줍니다. 

\begin{figure}[htbp]
    \centering
    \includegraphics[scale=0.5]{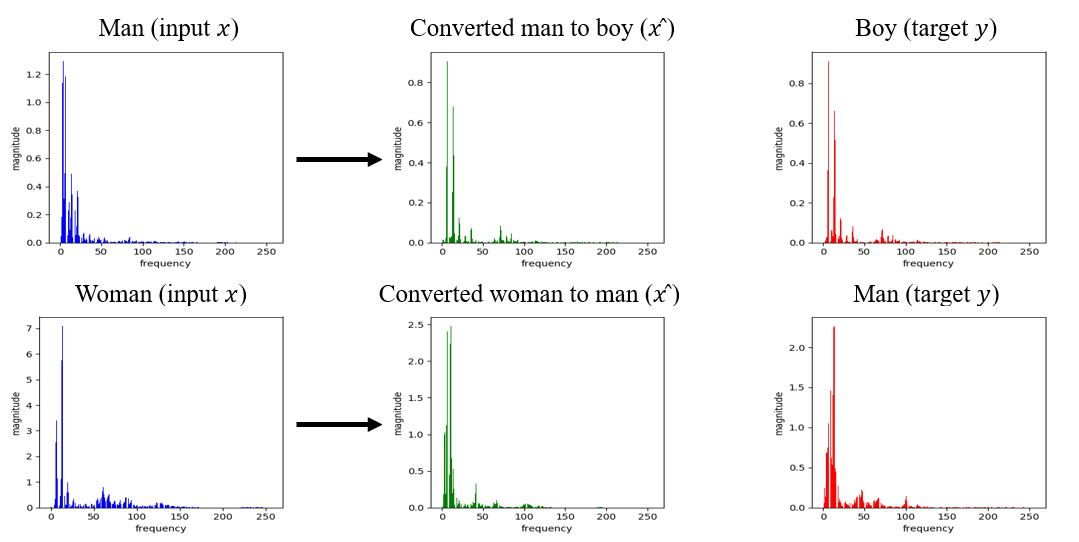}
    \caption{Visualization of our model conversion results about frequency difference. Same results of Fig.\ref{fig:spectrogram_results}. The x-axis consists of 256 cells because our input spectrum is used to 256 dimensions.}
    \label{fig:frequency_analysis}
\end{figure}

Fig.~\ref{fig:frequency_analysis} shows more accurate analysis of our conversion results. Each figure on the first row is the frequency with analyzed spectrum of man, result which is converted from man to boy, and spectrum of boy, respectively. The maximum y-axis in $man_{x}$ is near 1.3, in $boy_{y}$ is near 0.9, and in $\hat{x}$ which is our converted result is near 0.9. Frequency of $man_{x}$ in low-frequency cell is more higher than the frequency of $boy_{y}$. Through these analysis, it shows that low-frequencies from magnitude of $man_{x}$ are densely distributed and more higher than magnitude of $boy_{y}$. From the performance of model, the highest magnitude value in low-frequencies from $\hat{x}$ is near 0.8. This shows very close to the $boy_{y}$, and the $\hat{x}$ frequency distribution also similar to $boy_{y}$.

Likewise, each figure on the second row is the frequency which analyzed spectrum of woman, result which is converted from woman to man, and spectrum of man. The maximum y-axis in $woman_{x}$ is around 7.0, in $man_{y}$ is around 2.4, and our conversion result $\hat{x}$ is around 2.4. Before conversion, the highest magnitude in low-frequency cell is 7.0. However, the highest magnitude in $\hat{x}$ is around 2.4. Moreover, magnitude on $\hat{x_2}$ between 50 to 100 cell is cutting and closer to target $man_{y2}$ magnitude. In addition, the highest magnitude value on $woman_{x2}$ is near the seventh around $20th$ frequency cell. After the conversion, however, the highest magnitude value on $\hat{x_2}$ is under 2.5 around $20th$ frequency cell. Therefore the results, as shown in Fig. \ref{fig:frequency_analysis} indicate that our proposed model successfully performed the conversion.

%더 정확한 분석을 위해 Fig.\ref{fig:frequency analysis}의 first row의 왼쪽에 있는 그림은 man의 input이며, target인 boy의 frequency보다 더 높은 magnitude 값을 갖으며 low-frequency cell 부분에서 boy보다 촘촘하게 분포되어 있는 것을 볼 수 있습니다. 변환 후인 $\hat{x}$의 최고 magnitude 값은 target인 boy와 거의 근사해졌으며, 주파수 분포 또한 닮아간 것을 볼 수 있습니다. Second row의 그림들 역시 마찬가지로, woman의 가장 큰 magnitude는 20번째의 frequency dimension인 7 근처의 값입니다. 하지만 변환 후인 $\hat{x}$는 target과 같이 3이상이 되는 magnitude가 없는 결과를 보여줍니다.
%To get quantitative performance, we scored Naturalness, Similarity, and Clarity through MOS evaluation of 30 people.
To get quantitative performance, we randomly gather 38 adults from 20 to 30 ages. We measure our proposed model results using the metrics such as naturalness, similarity, and clarity with MOS, respectively. Sample of voices are randomly selected and same batches of samples are given to participants. Four sources, targets, and result samples of our model are extracted. Totally, 144 samples are evaluated.~\footnote{Audio samples are available at \url{https://kaen2891.github.io/e2e_vc_transformer_results/}} Source speakers and target speakers are different. 

\begin{table}[htbp]
\caption{MOS evaluations for naturalness with converted speech with 95\% confidence interval. Higher is more natural voice as human (1-5).}
\begin{center}
\scalebox{0.9}{
\begin{tabular}{|l|c|c|c|c|}
\hline
\diagbox[]{Source}{Target}& Man & Woman & Boy & Girl \\
\cline{1-5}
\hline
Man & - & 3.28$\pm0.29$ & 4.20$\pm0.53$ & 3.72$\pm0.33$ \\
\cline{1-5}
\hline
Woman & 2.82$\pm0.29$ & - & 3.18$\pm0.30$ & 3.45$\pm0.29$ \\
\cline{1-5}
\hline
Boy & 2.97$\pm0.31$ & 3.24$\pm0.27$ & - & 3.80$\pm0.28$ \\
\cline{1-5}
\hline
Girl & 3.01$\pm0.29$ & 3.56$\pm0.25$ & 3.53$\pm0.32$ & - \\
\cline{1-5}
\hline
\end{tabular}}
\label{mos_naturalness}
\end{center}
\end{table}
Table \ref{mos_naturalness} is an evaluation about how natural the converted voice is as human. We got the highest score ($4.20\pm0.53$) from conversion tasks from man to boy and the lowest score ($2.82\pm0.29$) from conversion tasks from woman to man. 
%TABLE \ref{mos_naturalness}은 변환된 음성이 얼마나 사람처럼 자연스러운지에 대한 평가입니다. Men to Boys conversion task에서 가장 높은 점수를 획득하였고 women to men task에서 가장 낮은 점수를 획득하였습니다. 
\begin{table}[htbp]
\caption{Similarity evaluation for the converted speech with 95\% confidence interval. Higher score is more similar to target voice (1-5).}
\begin{center}
\scalebox{0.9}{
\begin{tabular}{|l|c|c|c|c|}
\hline
\diagbox[]{Source}{Target}& Man & Woman & Boy & Girl \\
\cline{1-5}
\hline
Man & - & 3.91$\pm0.24$ & 4.36$\pm0.19$ & 4.26$\pm0.22$ \\
\cline{1-5}
\hline
Woman & 3.09$\pm0.31$ & - & 3.69$\pm0.31$ & 3.93$\pm0.24$ \\
\cline{1-5}
\hline
Boy & 3.30$\pm0.30$ & 3.50$\pm0.27$ & - & 4.28$\pm0.18$ \\
\cline{1-5}
\hline
Girl & 3.39$\pm0.30$ & 4.04$\pm0.20$ & 4.13$\pm0.24$ & - \\
\cline{1-5}
\hline
\end{tabular}}
\label{mos_similarity}
\end{center}
\end{table}
Table \ref{mos_similarity} is an evaluation about how similar the converted voice is to the target voice. Source speakers and target speakers are different. We got the highest similarity ($4.26\pm0.22$) from conversion tasks from man to boy and the lowest similarity ($3.09\pm0.31$) from conversion tasks from woman to man. 
%TABLE \ref{mos_similarity}는 변환된 음성이 대상 음성과 얼마나 유사한지에 대한 평가 부분입니다. 평균적으로 Naturalness 보다 높은 성능을 도출하였습니다. Men to boys conversion task에서 가장 높은 유사도를 보였고, women to men conversion task에서 가장 낮은 점수를 획득하였습니다.

\begin{table}[htbp]
\caption{Clarity evaluation for the converted speech with 95\% confidence interval. Higher score is more clear to script (1-5).}
\begin{center}
\scalebox{0.9}{
\begin{tabular}{|l|c|c|c|c|}
\hline
\diagbox[]{Source}{Target}& Man & Woman & Boy & Girl \\
\cline{1-5}
\hline
Man & - & 3.78$\pm0.27$ & 4.31$\pm0.19$ & 4.22$\pm0.21$ \\
\cline{1-5}
\hline
Woman & 3.57$\pm0.30$ & - & 3.83$\pm0.26$ & 3.80$\pm0.22$ \\
\cline{1-5}
\hline
Boy & 3.47$\pm0.26$ & 3.80$\pm0.26$ & - & 4.24$\pm0.22$ \\
\cline{1-5}
\hline
Girl & 3.84$\pm0.30$ & 4.00$\pm0.23$ & 4.24$\pm0.22$ & - \\
\cline{1-5}
\hline
\end{tabular}}
\label{mos_clarity}
\end{center}
\end{table}
Table \ref{mos_clarity} is an evaluation about how clear the pronunciation of converted voice to given script is. We get the highest clarity ($4.31\pm0.19$) from conversion tasks from man to boy and the lowest clarity ($3.47\pm0.26$) from conversion tasks from boy to man. The score of converting to children is high when they are targeted.
%TABLE \ref{mos_clarity}는 변환된 음성이 얼마나 script에 clear하게 들리는지에 대한 평가 부분입니다. Boy가 target일 때의 voice conversion에서 평균적으로 가장 높은 점수를 획득하였습니다. 또한 men이 target일 때의 voice conversion에서 평균적으로 가장 낮은 점수를 획득하였습니다.

In the overall speaker average MOS, the scores of our experiment results are $3.40\pm0.31$ in naturalness, $3.82\pm0.25$ in similarity, and $3.93\pm0.25$ in clarity, respectively. Our results show that the proposed method transforms voice with good clarity while maintaining appropriate naturalness and similarity.
%3개의 평가 test 중 naturalness는 평균 3.40$\pm0.31$을 similarity는 평균 3.82$\pm0.25$를, clarity는 평균 3.93$\pm0.25$를 각기 받았습니다. 우리의 결과는 제안 된 방법이 자연스럽고 유사성의 적절성을 유지하면서 좋은 선명도로 변환 할 수 있음을 보여 주었다.

\section{Conclusion}
\subsection{Summary}
We proposed a voice transform with self-attention mechanism in a raw spectrum level, while conventional methods use a vocoder in MFB level. MFB-based approaches had the advantage of computational learning convenience compared to raw spectrum. However, speech generator with MFB approaches require vocoder that needs a huge amount of computation expense for training process. With vocoder, it is possible to get better quality of the voice with the synthesis. On the contrary, the problem of complexity due to the extra computation is inevitable. The additional pre/post processing such as MFB and vocoder is not essential to convert real human speech to others. In this paper, we proposed a vocoder-free end-to-end voice conversion method using a fast and efficient transformer network that can convert spectrum in parallel manner. Obtaining the conversion results with raw spectrum without the help of repetitive vocoder had the advantage of using an original phase information to provide the result. We gathered 38 participants and conducted MOS evaluation on the naturalness, similarity and clarity of the converted speech. In the overall speaker average MOS, the scores of our experiment results got $3.40\pm0.31$ in naturalness, $3.82\pm0.25$ in similarity, and $3.93\pm0.25$ in clarity, respectively. Our results showed that the proposed method is possible to transform with good clarity while it is maintaining appropriateness of naturalness and similarity.

\subsection{Future Work}
In the evaluation phase, there was an unnatural converted part of $\hat{x}$. It seems to be caused by misalignments since the lengths of $\hat{x}$ and $phase_x$ deviate significantly. This is a feature of the transformer-based model which is converting to the maximum length. In other words, the length of all dataset are same because of zero-padding. However, if the actual vector length of $phase_x$ is less than the $\hat{x}$, it causes a serious problem that make misalignment. In the above case, the quality of the recovered waveform can be poor. Thus, the pitch is broken and the naturalness is weakened. Therefore, Our model need to phase transform to solve misalignment problem. The finding is unexpected and it suggests that there is problem related with the input spectrum length.

%평가 단계에서 $ \ hat {x} $의 부 자연스러운 변환 부분이있었습니다. $ \ hat {x} $와 $ phase_x $의 길이가 크게 다르기 때문에 정렬 불량으로 인한 것으로 간주됩니다. 이는 트랜스포머 기반 모델이 최대 길이로 변환하기 때문입니다. 다시 말해, 모든 데이터 세트의 길이는 제로 패딩으로 인해 동일합니다. 그러나 $ phase_x $의 실제 벡터 길이가 $ \ hat {x} $보다 작 으면 정렬 불량이 발생하는 심각한 문제가 발생합니다. 위의 경우 복구 된 WAV의 품질이 떨어질 수 있습니다. 따라서 피치가 깨지고 자연 스러움이 약해집니다. 따라서 우리의 모델은 오정렬 문제를 해결하기 위해 위상 변환이 필요합니다. 이 결과는 예상치 못한 결과이며 입력 스펙트럼 길이와 관련된 문제가 있음을 나타냅니다.

We found the importance of $phase$ in the study. The problem can be solved if $phase_x$ and the converted $\hat{x}$ are aligned with each other. To solve the problem, we have to use complex neural network\cite{trabelsi2017deep} to align the magnitude and phase which are included in the raw spectrum. If the phase could be aligned based on the converted magnitude, the quality of human voice will be improved. It will be possible to convert voices of minorities with poor speech recognition performance to those of common adults.  It is available to achieve better speech recognition performance through speaker adaptation which replaces the features of minorities and common adults. We are going to research phase adaptation and alignment with magnitude as our next task.
%이 연구에서 $phase$의 중요성을 발견했습니다. $ phase_x $와 변환 된 $ \ hat {x} $가 서로 정렬되면 이 문제를 해결할 수 있습니다. Complex Neural Network를 사용하여 raw spectrum과 magnitude, 및 phase 간의 관계를 조정하여 magnitude와 위상 사이의 변화를 맞추면 사람의 목소리 모양이 더 부드러워지고 품질이 향상 될 것입니다. 따라서, 음성 인식 성능이 좋지 않은 소수자들의 음성을 일반 성인의 음성으로 변환 할 수 있습니다. 스피커 적응을 위해 소수자와 일반적인 성인들의 특징을 대체하여 더 나은 음성 인식 성능을 달성 할 수 있습니다. 우리는 이러한 영역으로 확장하기 위해 미래의 연구를 위해 위상 적응과 위상 정렬을 규모에 맞게 조준합니다.
\section*{Acknowledgment}
This work was supported by Institute of Information \& Communications Technology Planning \& Evaluation(IITP) grant funded by the Korea government(MSIT) (2016-0-00564, Development of Intelligent Interaction Technology Based on Context Awareness and Human Intention Understanding)

\bibliographystyle{IEEEtran}
\bibliography{IEEEexample.bib}
\end{document}